\documentclass{kluwer}    
\usepackage{graphicx}

\newdisplay{guess}{Conjecture}

\begin{document}

\begin{article}
\begin{opening}         
\title{The Global, Local and Cluster Galaxy Luminosity Function} 
\author{Simon Driver and Roberto De Propris}  
\runningauthor{Simon Driver and Roberto De Propris}
\runningtitle{The Global Luminosity Function}
\institute{Research School of Astronomy and Astrophysics \\
Australian National University \\
Weston Creek, AUSTRALIA}
\date{Dec 1, 2002}

\begin{abstract}
We review selected measurements of the galaxy luminosity function
including the global field, the local group, the local sphere, nearby
clusters (Virgo, Coma and Fornax) and clusters in general. We conclude
that the overall cluster luminosity function is consistent with the
global luminosity function over the magnitude range in common ($-22
\leq M_B -5\mbox{log}h_{0.68} \leq -17 $). We find that only in the
core regions of clusters ($r \leq 300$ kpc) does the overall form of
the luminosity function show significant variation. However when the
luminosity function is subdivided by spectral type some further
variations are seen. We argue that these results imply: substantial
late infall, efficient star-formation suppression, and the confinement
of mass-changing evolutionary processes to the core regions only.
\end{abstract}

\end{opening}

\section{Introduction}
The Schechter luminosity function \cite{Schechter1976} has been the
standard expression for representing the space density of galaxies
over the past 25 years.  It has strengths and weaknesses. The
strengths are: its connection to the fundamental theory of the growth
of initial mass perturbations \cite{PS74}, its overall simplicity
(with three free parameters: the characteristic luminosity $L^*$; the
normalisation $\phi^*$; and the faint-end parameter $\alpha$), and its
simple analytical connection to the luminosity-density (and ultimately
via a mass-to-light ratio to the galaxian matter-densities).  Its
weaknesses are: the correlation of the three parameters, the critical
dependence of all three parameters upon the ``turn-over'' region, and
its inability to reflect deviations of the space-density from a simple
power-law at faint luminosities.

Over the past two decades numerous attempts have been made to
constrain the three defining Schechter function parameters in both
field and cluster environments producing a range of conclusions from a
ubiquitous luminosity function to strong environmental dependencies.
Here we showcase selected recent results from a variety of
methodologies, to address this specific question as to whether the
galaxy luminosity function is ubiquitous or not.

\begin{figure}
\includegraphics[height=7.5cm,width=12.0cm]{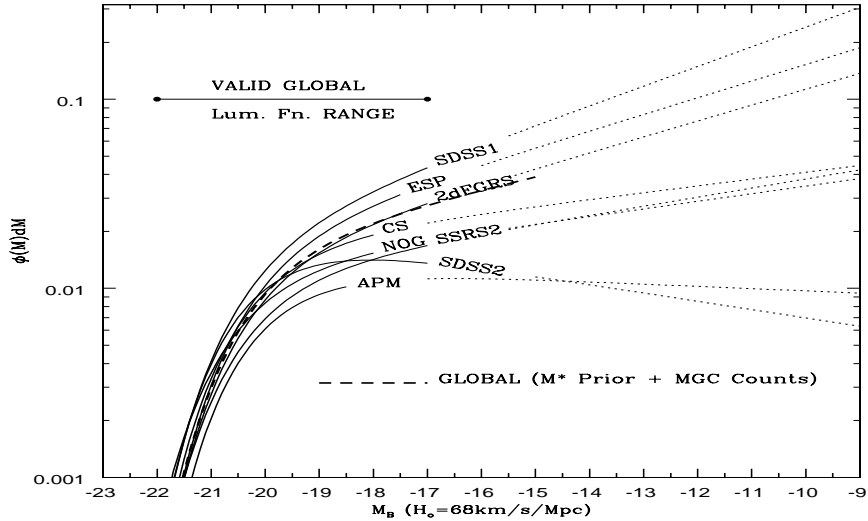}
\caption{The recently published blue-band Schechter functions transformed
to the standard Johnson B system. Surveys shown are:
APM (Loveday et al 1992);
ESP (Zucca et al 1997);
SSRS2 (Marzke et al 1998);
NOG (Marinoni et al 1999);
SDSS1 (Blanton et al 2001); 
CS (Brown et al 2001);
2dFGRS (Norberg et al 2002); 
SDSS2 (Blanton et al 2003).
}
\end{figure}

\section{The Global Luminosity Function}
So let us start with the field or global luminosity function.  Fig. 1
shows a compendium of eight recently derived $b, B, g$ or $V$-band
luminosity functions transformed to a common bandpass, $B$ (see Liske
et al 2003). Deriving mean values and errors from these data yield:
$M^*-5\mbox{log}h_{0.68} = -20.47 \pm 0.2$ mag, $\alpha=-1.1 \pm 0.1$,
$\phi^*=(0.006 \pm 0.001)h_{0.68}^3/$Mpc$^3$, providing a crude idea
of the systematic uncertainties (recall the values are strongly
correlated).  Clearly the precise global survey one adopts will be
critical for any field/cluster comparison.  So which survey to adopt
? This really depends upon the likely causes of these variations.
Most likely $\alpha$ is under-constrained due to insufficient
statistics at the faint-end, compounded by the luminosity-surface
brightness relation (Driver 1999). This can result in lower luminosity
galaxies being preferentially missed in shallower surveys (see Cross
\& Driver 2002). $M^*$ is presumably due to calibration errors and
$\phi^*$ to cosmic variance. Under these latter assumptions one can
adopt the two largest surveys (the 2dFGRS; Norberg et al 2002 and the
SDSS2; Blanton et al 2003) as the most credible. However although
these now agree at the bright end they show discrepancy at the
faint-end (see Fig. 1).

\begin{figure}
\includegraphics[height=7.5cm,width=12.0cm]{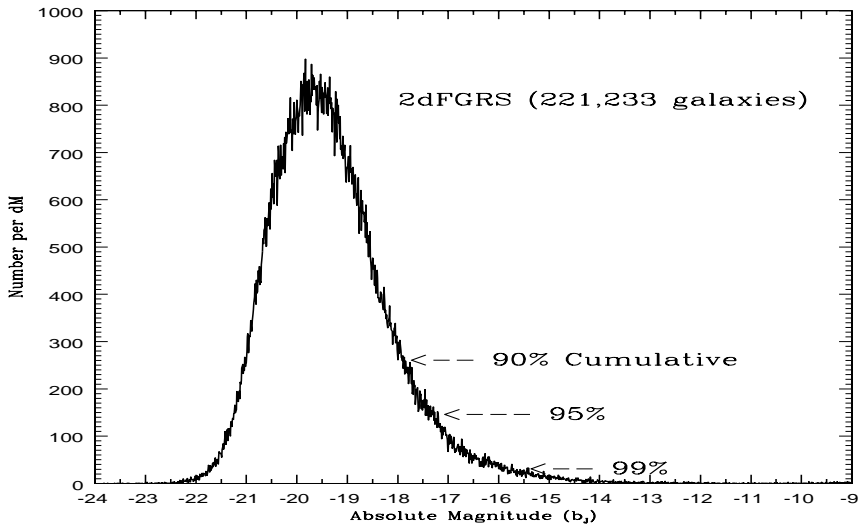}
\caption{The {\it observed} absolute magnitude distribution for the
two-degree field galaxy redshift survey. Given an incompleteness of
8\% we cannot reliably constrain the luminosity distribution fainter
than $M \approx -17$.}
\end{figure}

One extremely important point, particularly with respect to
comparisons with cluster LFs, is the very limited absolute magnitude
range over which the global LF is known. To highlight this Fig. 2
shows the {\it observed} distribution of the largest galaxy redshift
survey to date (the 2dFGRS; Colless et al 2001).  The 2dFGRS contains
221,233 galaxies of which only 9,398 galaxies are fainter than $0.1
L^*$ (see Fig. 2). With an incompleteness of 8\% the resulting
luminosity distribution can only be considered absolutely secure,
i.e. robust to any spectroscopic completeness bias, over a fairly
restricted magnitude range $-22 \leq M_{b_{j}}-5\mbox{log}h_{0.68}
\leq -17$. This inability to pin down the faint-end slope via a direct
approach severely restricts the range of comparison between the field
and cluster environments. That the two most recent and largest
redshift surveys show significant discrepancy in $\alpha$ is obviously
a cause for concern.

One entirely orthogonal way to constrain $\alpha$ which is {\it
robust} to most selection effects, has recently been demonstrated by
Liske et al (2003). They use the {\it curvature} of the bright
Millennium Galaxy Catalogue precision number-counts alone to constrain
$\phi^*$ and $\alpha$ by adopting a prior for the mean characteristic
luminosity from the latest redshift surveys
($M^*=-20.47+5\mbox{log}h_{0.68}$). The justification is that the
redshift surveys optimally sample the $L^*$ point and, with this well
defined, the curvature of the counts now depend entirely upon $\alpha$
(assuming the cosmology, k-corrections and evolutionary corrections
are known).  Fig. 3 shows the 1- 2- and 3-$\sigma$ confidence contours
in the $\phi^* - \alpha$-plane. The MGC results agree remarkably well
with the 2dFGRS (see Fig. 1) and we therefore adopt the 2dFGRS/MGC
parameters as our global yardstick (see Table 1).

\section{The Local Group and Cluster Luminosity Functions}
We now move to the luminosity functions of the near and dear.  Fig. 3
shows a compendium of the most recent local group, local sphere,
Virgo, Fornax and Coma luminosity functions (see Fig. 3 caption for
references). Where possible the most recent and largest datasets have
been selected to reflect a measure of the entire cluster LF.  In
particular we note that in all cases the group/cluster membership is
based on either direct distance indicators (local group and local
sphere), eyeball assessment (Virgo and Fornax) or spectroscopy (Coma)
as opposed to blind background subtraction. Also shown (Fig. 3 near
right and far right) are the 1-,2- and 3- $\sigma$ error-ellipses
derived from a simple $\chi^2$-minimisation of the Schechter
function. Fig. 3 (near right) are those derived from fitting the
entire available magnitude range and Fig. 3 (far right) are those
derived from fitting only over a luminosity range equivalent to the
known global LF range (see Figs 1 \& 2).  From Fig. 3 (far right) we
must conclude, contrary to some studies, that all environments are
consistent with each other and the global luminosity function. However
the analysis is limited by the insufficient number of members in the
few clusters studied compounded by the limited range over which the
global luminosity function is known.  Perhaps more compelling is the
apparent consensus between the overall luminosity functions from the
sparse local group to the dense Coma cluster.

\begin{figure}
\includegraphics[height=16.0cm,width=12.0cm]{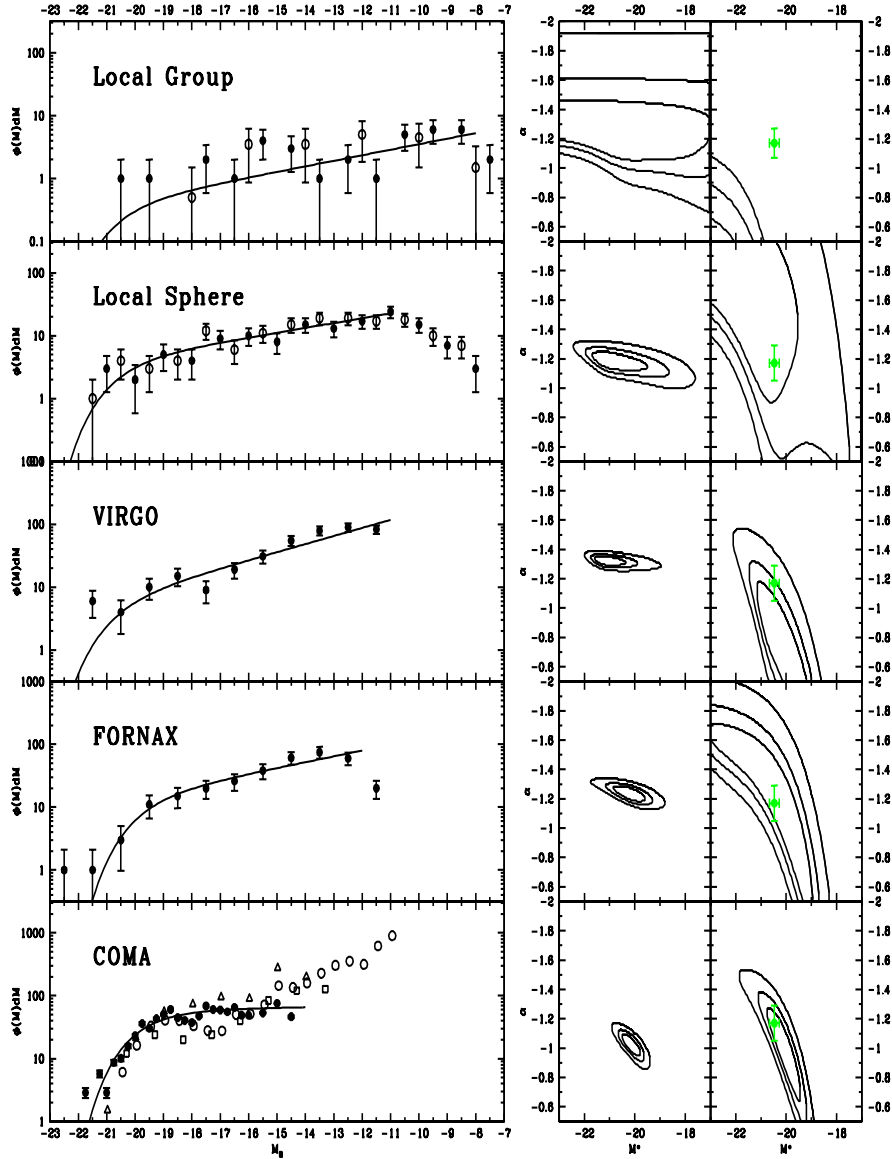}
\caption{The nearby group and cluster luminosity functions (left) and
$1-, 2-$ and $3-\sigma$ error ellipses for a Schechter function fit to
either the full luminosity range ($-22 < M <$ various, middle) or the
range comparable to the field range ($-22 < M -5\mbox{log}h < -17$,
right), all fits are to the solid data points.  Local group data:
Pritchet \& van den Bergh 1999 (solid); Mateo 1998 (open), Local
sphere data: Jerjen, Binggeli, Freeman 2000 (solid); Karachtenseva et al 2002
(open), Virgo data: Trentham \& Hodgkin 2001 (solid); Fornax data:
Ferguson 1989 + Deady et al 2003 (solid); Coma data: Mobasher et al
2003 (solid); Beijersbergen et al 2002 (open triangles), Trentham 1998
(open circles) and Andreon \& Culliandre 2002 (open squares).  The
solid line shows, in each case, the optimal Schechter function fit
over the full range of data. The data point and errorbars on the right
most panels shows the location of the adopted global luminosity
function.}
\end{figure}

\subsection{Coma}
The Coma cluster is worth some brief further comment as there have
been a number of recent studies with intriguing and even conflicting
results. On the largest scale Beijersbergen et al (2002) sample 5.2 sq
degrees using background subtraction to determine the overall LF
finding an anomalously low $M^*$ ($-19.0+5\mbox{log}h_{0.68}$) when
compared to the 1 sq degree spectroscopic study of Mobasher et al
(2003). Conversely Mobasher et al find a significantly flatter
faint-end slope than Beijersbergen et al.  More restricted surveys
such as Trentham (1998) and Andreon \& Culliandre (2002) which focus
on the core find evidence for an upturn and/or dip inconsistent with
the larger area studies (see Fig. 3).  However Beijersbergen et al show
that when they confine themselves to the Trentham region they also
recover the identical dip and upturn as seen by Trentham.  A plausible
explanation from a close reading of the relevant papers is the
following. Beijersbergen et al may have over estimated their field
counts at the bright-end, (see their Table 4 Column 6 which is
flatter than the expected $\log \mbox{ndm} \propto 0.6m$ relation at
bright mags), this will have minimal impact upon their core LF where
the cluster contrast is higher --- hence agreement with Trentham ---
but results in a significant bright-end over-subtraction of the full
cluster area --- hence disagreement with Mobasher et al.  Meanwhile
Mobasher et al may suffer from some B-band incompleteness in their
R-band selected spectroscopic sample.  Finally the Trentham and
Andreon \& Culliandre LFs are dominated by the core population and the
additional structures (LFs dips, upturns) must therefore be core
phenomena resulting from strong evolutionary processes in the core
region.  Clearly the definitive study of Coma remains to be done and
represents an excellent opportunity for a space-based survey where
high-resolution imaging can be used to determine accurate cluster
membership.

\section{Composite Cluster LFs}
For clusters beyond Coma the majority of studies have relied on the
background subtraction of the field galaxy population to recover the
cluster luminosity function to comparably faint absolute magnitudes
(e.g., Driver et al 1994).  As the discussion on Coma above suggests
the method of background subtraction, while efficient, is susceptible
to a variety of errors. In Driver et al (1998) we discuss the
criterion for successful background subtraction in substantial
detail. In essence the process is only reliable for very rich clusters
where the contrast of the cluster population is high. Valotto, Moore
\& Lambas (2001) more recently argue for a bias towards steeper
faint-end slopes because of preferential alignments between clusters
and filaments. While this argument has some merit locally it does not
hold for clusters at intermediate redshifts $z > 0.1$.  The reason for
this is that structures are correlated only over scales of order
100Mpc minimising the impact upon clusters at appreciable distances.
However Valotto et al do highlight the potential impact of cosmic
variance in general and some analysis have neglected to include this
significant error component\footnote{ For completeness we include a
prescription for calculating it from the angular 2pt correlation
function starting from Phillipps \& Disney (1985) and adopting
$A_W\theta^{-0.8}$ as the form of the background clustering, thus:

$\Delta N_{C}(m)^2 = N(m)^2\theta^4\omega(\frac{\sqrt{2}\theta}{3}) =
N(m)^{2} A_w(m)\theta^{3.2}(\frac{\sqrt{2}}{3})^{-0.8}$

\noindent
where $\Delta N_C(m)$ is the appropriate clustering error to be added
in quadrature to the Poisson error, $N(m)$ are the background galaxy
number counts per square degree, $\theta$ is the diameter of the
field-of-view and $A_w(m)$ is the amplitude of the angular correlation
function. From Metcalfe et al (1995) and Roche \& Eales (1999) we note
that: $N(m)=10^{0.377 (R-12.2)}$ and $A_w=10^{(-0.234 R+2.6)}$ for $20
< R < 25$.}.

Compounding this issue is also the ad hoc coverage of the various
studies.  In particular the majority of deep studies which have
recovered LFs to faint absolute magnitudes sample only the core region
identifying features such as dips and upturns similar to that seen in
Coma. Overall the conclusions fall into two opposing camps, those
that find significant variations between clusters (for example Garilli
et al 1999; Goto et al 2002 and qv) and those that do not (for
Paolillo et al 2001 and qv). The derived composite LFs for these three
surveys are shown in Table 1.  So are the variations real or an
artifact due to factors such as areal coverage, cosmic variance and/or
surface brightness incompleteness between the cluster and reference
fields ?  Until direct membership is ascertained via either
spectroscopy, photometric redshifts or high-resolution space-based
imaging it is unlikely that results from background subtraction can be
taken as conclusive either way.

\begin{figure}
\includegraphics[height=5.0cm,width=12.0cm]{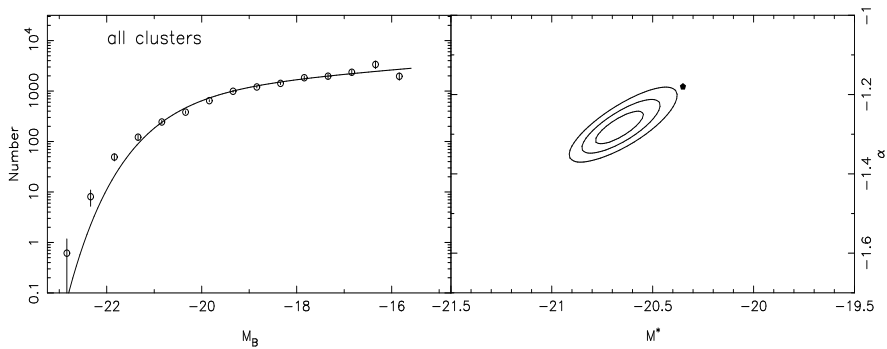}
\caption{The global (solid line) and composition cluster LF (open
circles) from within the 2dFGRS survey (left) along with the
appropriate errors (right).  Once again we find that the field and
cluster LFs are in statistical agreement.}
\end{figure}

\begin{figure}
\includegraphics[height=8.0cm,width=12.0cm]{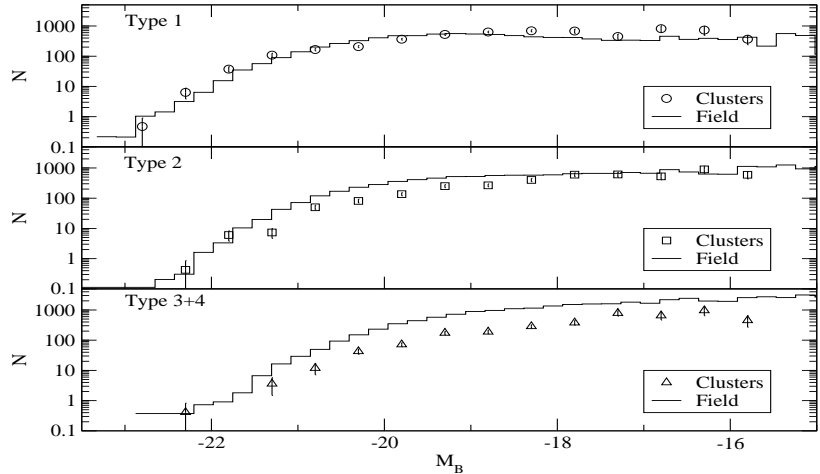}
\caption{The global (solid lines) and composition cluster LF (open
circles) subdivided according to spectral type from early (top) to
intermediate (middle) and late (bottom).}
\end{figure}

\section{The 2dFGRS Cluster and Field Luminosity Functions}
From the above it is clear that, for the moment, the only reliable way
to fully assess the environmental impact on the LF is from within a
single large spectroscopically confirmed and self-consistent
catalogue. The 2dFGRS offers such an opportunity (see De Propris et al
2003). The 2dFGRS is a magnitude limited redshift survey to an
approximate limit of $b_{j} = 19.45$ with 221233 redshifts and 92\%
completeness. The survey has median redshift of $\sim 0.12$ and
contains 60 well sampled nearby rich clusters (see De Propris et
al. 2002). Full details of the analysis are given in De Propris et al
(2003) however Fig. 4 shows the end result comparing the derived field
(solid line) and composite cluster (open circles) LFs along with the
error ellipses. Note that the field and composite cluster LFs are
based upon $\sim75,000$ and 4,186 spectroscopically confirmed galaxies
respectively.  The formal Schechter function fits are also shown in
Table 1. These suggest a marginally brighter $M^*$ and marginally
steeper $\alpha$.  Perhaps more important though De Propris et al go
on to subdivide their sample of 60 clusters according to; rich versus
poor, high versus low velocity dispersion, Bautz-Morgan class
I,I-II,II versus II-III,III and inner versus outer. In all cases the
composite LFs remain consistent with the original composite LF,
although it is noted that the composite {\it inner} LF is poorly fit
by a Schechter function indicating an upturn comparable to that seen
in the core Coma studies.  Statistically though the only significant
variation comes when the field and cluster LFs are subdivided
according to type. In these case it is seen that the early-type
galaxies have both a slightly brighter and steeper LF while the
late-type galaxies remain unchanged in shape but reduced in terms of
relative normalisations, see Fig. 5.

\section{Conclusions}
At face value we must conclude that the galaxy luminosity function
appears ubiquitous across all environments studied with only marginal
evidence for a slightly brighter $M^*$ and a slightly steeper $\alpha$
in clusters (see also the article in these proceedings by Christlein).
Only in the very core regions of rich clusters or when the population
is sub-divided according to spectral type are significant variations
seen.  We postulate that these results lead to three likely
conclusions:

\noindent
(1) Cluster infall is an ongoing process with a substantial fraction
of the cluster population infalling in recent times. This explains the
universality of the LF from the field to rich clusters and constrains
any epoch of major merging to have occurred prior to the epoch of
cluster formation.

\noindent
(2) Star-formation is effectively and efficiently inhibited in the
infalling population however the halo merger rate is low. This
explains the variation with spectral type but not in the overall
LF. Hence while galaxies may have shifted their spectral classes their
broad-band output remains mostly unchanged.

\noindent
(3) Dramatic evolutionary processes (merger, harassment etc) resulting
in the construction of the cD/D galaxies and destruction of the dwarf
population is confined to the core region. This explains the dips and
upturns seen in Coma, the 2dFGRS core sample and numerous deep
background subtracted core studies.

~

\noindent
This is a data rich time for this field with the development of wide
field mosaics, the Advanced Camera for Surveys and the upcoming
large-format infrared detectors. We look forward to the power of these
technologies being brought to bear on this unfolding story.

\begin{table}[h]
\caption{A summary of Schechter LF estimates.}
\begin{tabular}{ccll} \hline
Survey     & Limit   & $M_{B}^{*}-5\mbox{log}h_{0.68}$ & $\alpha$ \\ \hline
{\bf GLOBAL}         & $M_B \leq -17$ & $-20.47 \pm 0.2$ & $-1.17 \pm 0.02$ \\
{\bf LOCAL GROUP}    & $M_B \leq  -7$ & $-21^{+4}_{-2}$        & $-1.2^{+0.24}_{-0.14}$ \\
{\bf LOCAL SPHERE}   & $M_B \leq -11$ & $-20.1^{+1.2}_{-0.6}$  & $-1.19^{+0.05}_{-0.06}$ \\
{\bf VIRGO}          & $M_B \leq -11$ & $-20.2^{+0.7}_{-0.5}$  & $-1.32^{+0.03}_{-0.03}$ \\
{\bf FORNAX}         & $M_B \leq -12$ & $-20.2^{+0.5}_{-0.4}$  & $-1.23^{+0.04}_{-0.03}$ \\
{\bf COMA}           & $M_B \leq -14$ & $-20.1^{+0.25}_{-0.20}$  & $-1.01^{+0.04}_{-0.05}$ \\
{\bf Garilli et al}$^1$  & $M_{B} \leq -17$ & $-20.6 \pm 0.1$  & $-0.97^{+0.04}_{-0.02}$ \\
{\bf Paolillo et al}$^1$ & $M_{B} \leq -16$ & $-20.5^{+0.13}_{-0.17}$ & $-1.07^{+0.09}_{-0.07}$ \\
{\bf Goto et al}$^1$     & $M_{B} \leq -16.5$ & $-21.47 \pm 0.11$ & $-1.00 \pm 0.06$ \\
{\bf 2dFGRS FIELD}   & $M_B \leq -17$ & $-20.35 \pm 0.04$  & $-1.19 \pm 0.01$ \\
{\bf 2dFGRS CLUSTER} & $M_B \leq -17$ & $-20.62 \pm 0.07$  & $-1.28 \pm 0.03$ \\ \hline
\end{tabular}

\noindent
$^1$ $g$-band data converted to $B$ assuming $B=g+0.54$

\vspace{-1.0cm}

\end{table}

\vspace{-0.3cm}

\acknowledgements 

\vspace{-0.3cm}

We acknowledge invaluable access to the two-degree
field galaxy redshift survey database and note that the phenomenal
success of the 2dFGRS is largely based on the hard work and dedication
of the staff of the Anglo-Australian Observatory.  We thank the
organisers of the JENAM2002 Galaxy Evolution Workshop for a very enjoyable
meeting.

\end{article}
\end{document}